\documentclass[]{article}
\usepackage{graphicx}
\usepackage{amssymb}
\usepackage[numbers]{natbib} 
\usepackage{xcolor}

\title{U.S. Social Fragmentation at Multiple Scales}

\author
{Leila Hedayatifar, Rachel A. Rigg, Yaneer Bar-Yam and Alfredo J. Morales$^\ast$ \\
\normalsize{New England Complex Systems Institute}\\
\normalsize{277 Broadway, Cambridge, MA, 02139}\\
\normalsize{$^\ast$ E-mail: alfredo@necsi.edu}
}
\date{}

\begin{document}

\maketitle

\begin{abstract}
Despite global connectivity, societies seem to be increasingly polarized and fragmented. This phenomenon is rooted in the underlying complex structure and dynamics of social systems. Far from homogeneously mixing or adopting conforming views, individuals self-organize into groups at multiple scales, ranging from families up to cities and cultures. In this paper, we study the fragmented structure of the American society using mobility and communication networks obtained from geo-located social media data. We find self-organized patches with clear geographical borders that are consistent between physical and virtual spaces. The patches have multi-scale structure ranging from parts of a city up to the entire nation. Their significance is reflected in distinct patterns of collective interests and conversations. Finally, we explain the patch emergence by a model of network growth that combines mechanisms of geographical distance gravity, preferential attachment, and spatial growth. Our observations are consistent with the emergence of social groups whose separated association and communication reinforce distinct identities. Rather than eliminating borders, the virtual space reproduces them as people mirror their offline lives online. Understanding the mechanisms driving the emergence of fragmentation in hyper-connected social systems is imperative in the age of the Internet and globalization.

\end{abstract}

\section{Introduction}\label{intro}
The increasing polarization of societies is becoming apparent around the world. Despite access to global communication \cite{I18}, people seem to be splitting into groups that mostly listen to their own members \cite{m11,I3,I28}. Individual choices of association due to ideologies \cite{f3,I2,I8}, occupations \cite{I1,I29}, or consumer habits \cite{m5} can drive the emergence of social polarization or fragmentation \cite{I23,I29}. While different social features affect processes of homophily and influence, in this work we study how fundamental geographical factors also affect the large-scale structure of social interactions and communication networks. Previous studies have proposed distance as the driving factor for social interactions \cite{Liben-Nowell11623,m9, m2}. We show that the structure of the emergent social networks is richer than what distance alone can explain and includes the influence of factors like administrative borders and urban structures. It is crucial to understand the structural and geographical properties of collective association and their relationship to the social space. 

The social space is defined as the place where people meet and interact \cite{I15}. While group cohesion is strongly influenced by internal communication, weaker external ties are necessary for integration at larger scales, providing individuals with information and resources beyond the borders of their own community \cite{I15, I13,r1,r2,r3, Kleinberg2000}. Previous studies have shown that the structure of both strong and weak ties affects the behavior of social systems, including the spread of innovation \cite{I16}, business and culture \cite{I17}, crime systems \cite{r4}, and the development of regional and national events \cite{I14}. Social fragmentation affects the way information flows among individuals \cite{I12} and consequently their emergent behaviors \cite{f3,I25,I29}, including political or physical conflict \cite{I30,I32,I33,I34}. 

The recent availability of large-scale datasets obtained from communication or transaction records for landlines, mobile phones, social media, and banknote circulation has considerably improved our ability to study social systems \cite{I24, I20, Gonzalez2008, m12}. Geo-located data sources, such as Twitter, enable direct observation of social interactions and collective behaviors with unprecedented detail. While the Twitter user base is known to skew younger and more urban \cite{Smith2018, Duggan2015}, the large size of its user base and high frequency of tweets has enabled new types of studies of networks and geo-located activities. For example, Twitter data has been utilized in studies on a wide range of behavioral phenomena, including human migration, disease outbreaks, and patterns of happiness and lifestyle \cite{Kallus2015, Sitko2014, Frank2013a, Blanford2015,Bakillah2015, Junjun2017}.

Networks of human mobility \cite{I24,m12,I21,I5,I22} and communication \cite{m5,m12,I4,I6,I7,I9,Kallus2015} reveal the existence of geo-located communities or patches. Researchers have used Twitter data on mobility to show where geo-located communities deviate from administrative boundaries in Great Britain \cite{Junjun2017}. Others have generated networks of Twitter communications and examined community formation in various countries \cite{Kallus2015} or in a natural disaster \cite{Bakillah2015}. While these studies analyze the structure of mobility or communication networks separately, we show that these two are not independent from one another and rather that networks in physical space are mirrored in the virtual space.

In this work, we utilize geo-located Twitter data to identify two networks in the U.S., human mobility and communication. We show that the specific geographic patches of both networks are very similar. We validate the significance of these patches by analyzing hashtag use by location and find similar patterns of divergence as in the mobility and communication networks. Finally, we build a model of network growth to understand the generic statistical properties of the natural human dynamics observed in the data. Our model combines a distance gravity component for cluster formation with preferential attachment and spatial growth mechanisms to allow clusters to differentiate in geographical space and grow over time. This work provides a comprehensive depiction of network dynamics and social fragmentation in the U.S.

\section{Materials and Methods}\label{material-methods}

\subsection{Data}\label{data} 
We use geo-located Twitter data to generate geographical networks based on where people travel or communicate. The data were obtained using the Twitter Streaming Application Programming Interface (API). We collected tweets from August 22, 2013 to December 25, 2013, totaling over 87 million tweets posted by over 2.8 million users in the U.S. 

\subsection{Networks}\label{networks} 

We analyze mobility and communication patterns by generating geographical networks. Nodes represent a lattice of  $0.1^{\circ}$ latitude $\times$ $0.1^{\circ}$ longitude cells overlaid on a map of the U.S. Each cell is approximately 10 km wide. There are about 400,000 cells comprising inhabited areas of the U.S. Network edges reflect two types of data: mobility and communication. In the mobility network, edges are created when a user $u$ tweets consecutively from two locations, $i$ and $j$. In the communication network, edges are created when a user $u$ at location $i$ mentions another user $v$ that has most recently tweeted at location $j$. The weight of an edge represents the number of people who either travel or communicate between $i$ and $j$. These networks aggregate the heterogeneities of human activities in a large-scale representation of social collective behaviors \cite{I27}.

\subsection{Methods}\label{methods} 
The term network fragmentation is often used in the literature to describe the process of network dismantling \cite{Braunstein2016, Antulov2018}. In this work, we use the term ``social fragmentation" to represent the modular structure of a social system due to absence of links and nodes. This is in line with terminology from other works that employ community detection methods such as the Girvan-Newman method \cite{Girvan2002}.

{We analyze social fragmentation by applying the Louvain method \cite{f4} with modularity optimization \cite{m10} to the mobility and communication networks obtained from Twitter data. The Louvian algorithm starts by considering each node as a single community. Iteratively, nodes move to the neighboring communities and join them to maximize modularity ($M$). Modularity is a scalar value $-1<M<1$ that quantifies how distant the number of edges inside a community are from those of a random distribution. Negative modularities occur when nodes are assigned to the wrong communities, zero occurs when all the nodes are assigned to a single community, and higher values represent increasingly optimal partitions as the values get closer to $1$ \cite{Newman8577,Barab1}.}

To study communities at multiple scales, we use a generalized version of modularity \cite{f4} that includes a resolution parameter $\gamma$. In the conventional modularity equation, $\gamma = 1$ and the same weight is given to observed links and expected links from a randomized network. In the generalized form, $\gamma <1$ gives more weight to the observed links, which generates larger communities, while $\gamma >1$ puts more weight on the randomized term and generates smaller communities. Because it is a method with multiple maxima, we chose partitions that are robust to multiple runs of the algorithm.

We validate the significance of the patches by observing hashtag use. We create a matrix whose rows represent locations and columns represent hashtags. In order to observe collective behaviors, we consider only those hashtags that were posted at least 500 times and locations with at least 20 tweets. We apply the term frequency-inverse document frequency (TF-IDF) transformation \cite{baeza2011modern,manning2010introduction} to the matrices in order to normalize the hashtags (columns of the matrix). We then apply principal component analysis (PCA) \cite{tipping1999probabilistic} to the hashtag matrix and retrieve the top 100 components, and then apply t-distributed stochastic neighbor embedding (t-SNE) \cite{maaten2008visualizing,van2014accelerating} to the resulting PCA matrix.

\section{Results}

\subsection{Social Fragmentation} 

\begin{figure*}[htp!]
\centering
\includegraphics[clip,trim={1.5cm 1.5cm 1.5cm 1.5cm},width=.83\textwidth]{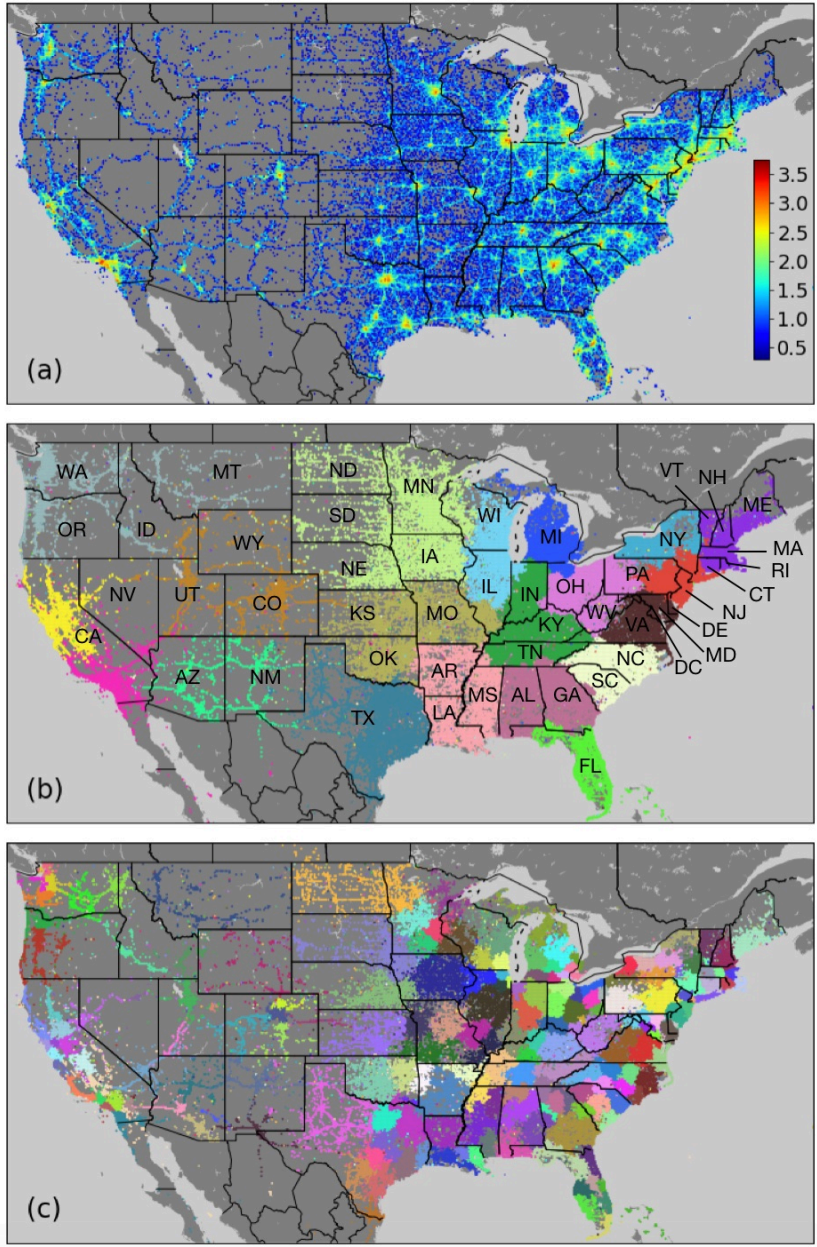}
\caption{\label{fig:Networks}%
Structure and fragmentation patterns of the network associated with human mobility. (a) Spatial degree centrality of the mobility network. Colors indicate the amount of people traveling at each location, measured by the logarithm of the degree centrality of each node (scale inset). The mobility network was used to generate communities using modularity optimization, with distinct colors indicating (b) 20 patches that can be visually associated to states or regions and (c) 206 smaller sub-communities within the communities of panel (b) that can be visually associated to urban centers.}
\end{figure*}

We first generated a mobility network of instances in which a user tweets from different locations, representing travel (see Section \ref{methods}). Figure \ref{fig:Networks} depicts the spatial properties of the mobility network on a map of the U.S. in terms of degree centrality (Figure \ref{fig:Networks} (a)) and two levels of modular structure (Figure \ref{fig:Networks} (b) and Figure \ref{fig:Networks} (c)). The degree centrality shows the density of user movements at each geographical point. The activity is concentrated in large cities (red in Figure \ref{fig:Networks} (a)) and decreases toward suburban and rural areas (green, blue and gray). In areas of the country with high population density, cities merge into large regions of high activity (e.g., the East Coast corridor). In other areas, roads are also visible, as people tweet when they travel between cities. Highways in rural areas with higher traffic appear in green, and less traveled roads are blue. 

The spatial fragmentation of social systems arises when people travel and choose which boundaries not to cross either directly or incidentally. Our results suggest that the U.S. mobility network is fragmented into $20$ large communities (Figure \ref{fig:Networks} (b)) whose boundaries often follow state boundaries but may in particular cases be parts of one state or the combination of multiple states. At a finer scale of subdivision, these large communities of the mobility network are subdivided into patches that typically include individual cities and their surrounding areas. There are $206$ such communities that we obtain by applying the same modularity optimization algorithm to each larger community (Figure \ref{fig:Networks} (c)).

\begin{figure*}[htp!]
\centering
\includegraphics[clip,trim={1.5cm 1.5cm 1.5cm 1.5cm},width=.83\textwidth]{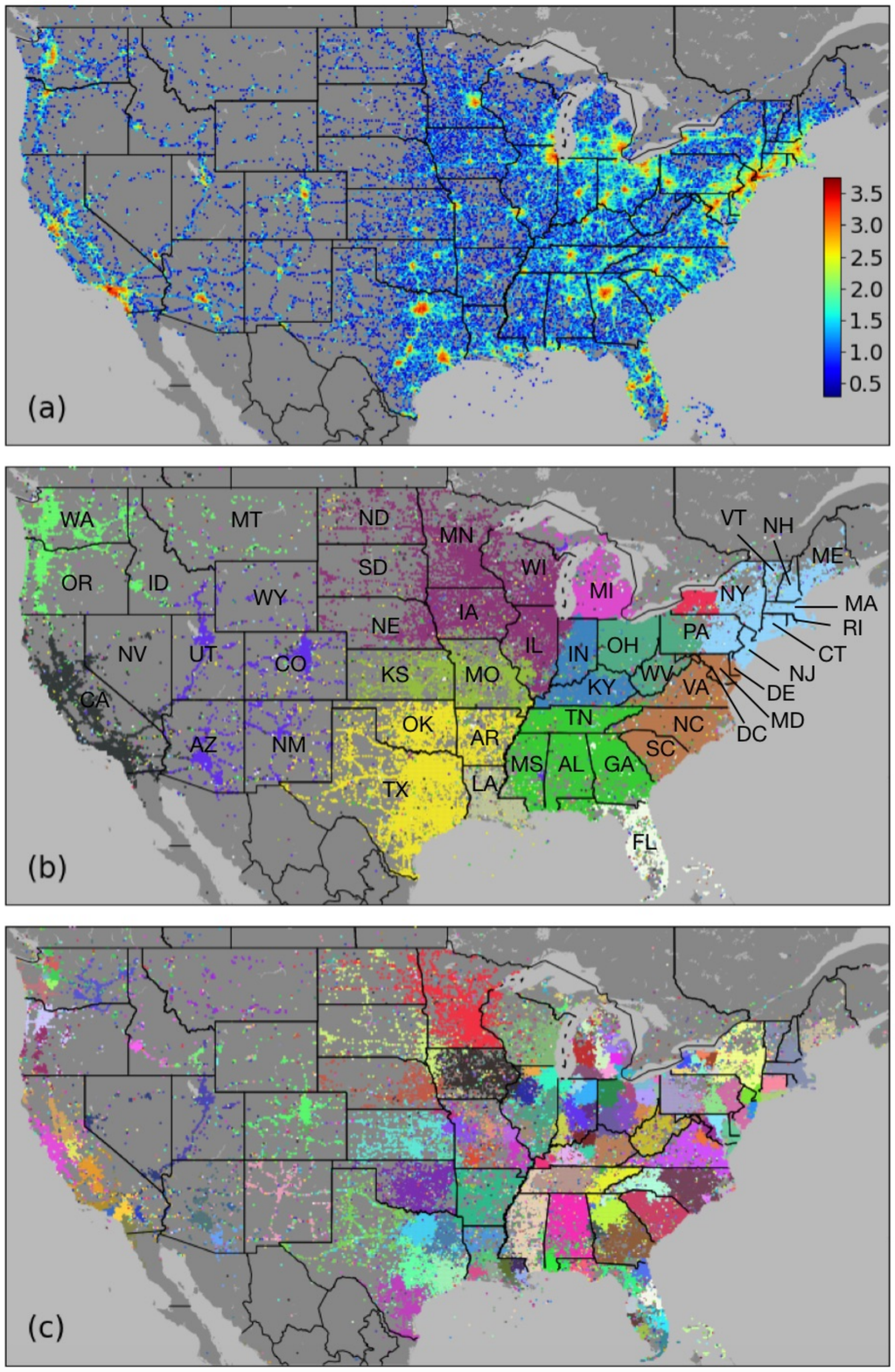}
\caption{\label{fig:NetworksS}
Structure and fragmentation patterns of the network associated with human communication. (a) Spatial degree centrality of the communication network. Colors indicate the amount of communication at each location, measured by  the logarithm of the degree centrality of each node (scale inset). The communication network was used to generate communities using modularity optimization, with distinct colors indicating (b) 15 patches that can be visually associated to states or regions and (c) 168 smaller sub-communities within the communities of panel (b) that can be visually associated to urban centers.}
\end{figure*}

Following the mobility network, we generated a communication network from Twitter mentions, shown in Figure \ref{fig:NetworksS}. Our modularity analysis on this network shows that it also has structure of social fragmentation that is consistent with the mobility network. Thus, while the Internet and social media have drastically affected the dynamics of communications, the geographic structure of online communication remains fragmented and presents a similar structure to the one obtained from offline interactions. There are some differences as well. In contrast to the 20 modules in the mobility network, there are 15 modules that arise in the communication network. 

The borders of some communities in Figure \ref{fig:NetworksS} are almost the same as those in the mobility network (Figure \ref{fig:Networks} (b)), such as the community encompassing states of the Northwest (WA, OR, ID, and MT), the community corresponding to Michigan (MI), and the community corresponding to Florida (FL). Ohio (OH), western Pennsylvania (PA) and West Virginia (WV) are also still in the same patch. Meanwhile, other communities in the mobility network merge into a larger community in the communication network. For example, the six-state region of New England (Maine (ME), Massachusetts (MA), New Hampshire (NH), Vermont (VT), Rhode Island (RI) and Connecticut (CT)) is a separate community in the mobility network but is combined with New York (NY), New Jersey (NJ), and Pennsylvania (PA) in the communications network. The two patches of North and South Carolina (NC and SC) and Virginia (VA) and Maryland (MD) are also combined into one. This demonstrates that certain areas have a broader radius of online communication than physical travel. Finally, Figure \ref{fig:NetworksS} (c) represents the smaller communities within each community in Figure  \ref{fig:NetworksS} (b). These patches show areas connected to urban centers and are very similar to those of the mobility network in Figure \ref{fig:Networks} (c). Some less populous states are now single communities, such as Montana (MT), Nebraska (NE), Kansas (KS), Oklahoma (OK), Arkansas (AR), and New Mexico (NM), while more densely populated areas are subdivided around urban centers.

To further investigate the role of state boundaries in community formation, we quantified to what extent each state contributes to communities for both networks (Section S1.1 and Figure S1). States mostly belong to specific communities. This shows that the structure we observe is not simply due to the effects of distance \cite{m9}. To show this, we generated artificial networks with links weighted by only the inverse of distance or distance squared (see Section S1.2). While spatial patches are also present in these artificial networks, the patches do not follow state boundaries and are not consistent across both types of networks (see Figures S2 and S3). We also performed validation of community stability and find that the number of communities and boundaries we show are consistent and stable across multiple realizations of the algorithm (see Section S1.3 and
Figure S4). Overlapping regions across realizations can happen either because small locations flip between large communities or because large communities are split into smaller ones.

\begin{figure}[t!]
\centering
\includegraphics[clip,trim={3.2cm 7.5cm 0.2cm 8.8cm},width=.8\textwidth]{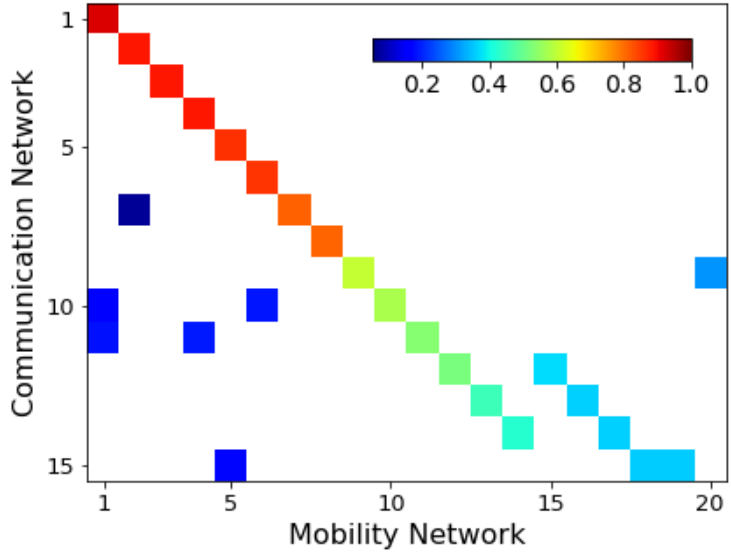}
\caption{\label{fig:Sim} Similarity of communities in the communication and mobility networks. Matrix of the regional communities for the communication network ($y$\mbox{-}axis, $n=15$) and mobility network ($x$\mbox{-}axis, $n=20$), ordered by decreasing overlap between communities. Cell colors represent the number of nodes overlapping between the two networks in each community, normalized by the size of the communities per row (scale inset), with no overlap indicated in white. }
\end{figure}

We quantitatively compared the modular structure of the mobility and communication networks (Figure \ref{fig:Sim}) by creating a matrix where we count the number of overlapping nodes of communities arising from the networks of communication (rows) and mobility (columns). Rows have been normalized by the size of each community in the communication network. Some communities from the communication network are almost identical in the mobility network and therefore show a high overlap (red). Others are similar but not identical. A few communities from the mobility network are merged into communities in the communication network (green and light blue). Despite the observed differences in the networks representing two fundamentally different types of interactions, the modular structure is remarkably consistent, revealing that there is a strong coupling between the way people travel in physical space and communicate with each other online. 

\begin{figure*}[htp!]
\centering
\includegraphics[clip,trim={0.1cm 2.5cm 0.1cm 2cm}, width=1\textwidth]{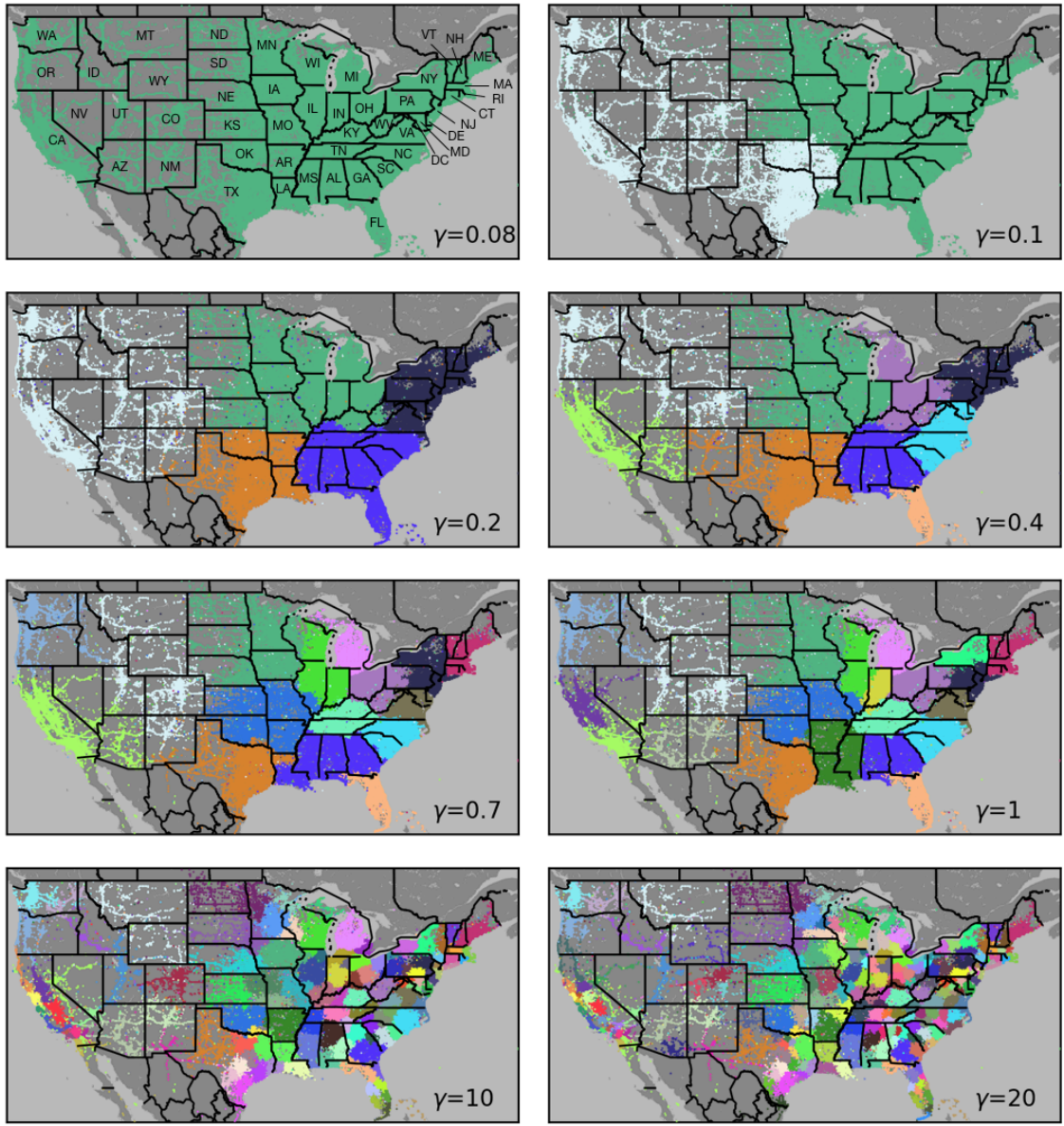}
\caption{\label{fig:Mob} Multi-scale decomposition of the mobility network. Colors indicate geographical patches detected in the mobility network for values of the resolution parameter $\gamma$ varied from $0.08-20$ (upper left to bottom right). Colors are retained across panels by the following rule: when a community is divided into multiple sub-communities, the sub-community that is the most connected to the original (``parent") community retains the color of the parent community; the other sub-communities are assigned new colors. The modularity for all of the panels is over 0.8.}
\end{figure*}

\begin{figure*}[htp!]
\centering
\includegraphics[clip,trim={0.1cm 2.5cm 0.1cm 2cm}, width=1\textwidth]{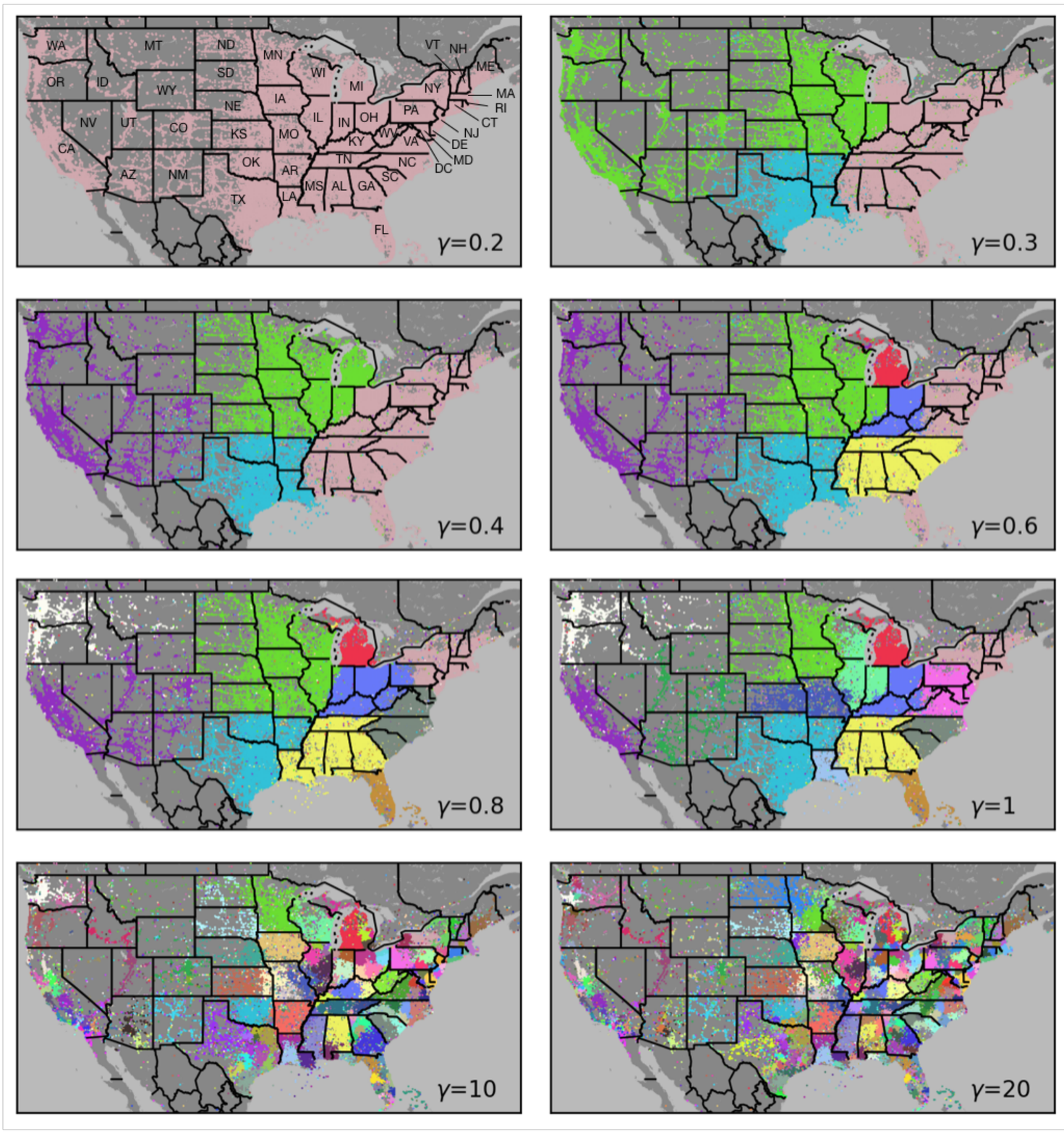}
\caption{\label{fig:Men} Multi-scale decomposition of the communication network. Colors indicate geographical patches detected in the communication network for values of the resolution parameter $\gamma$ varied from $0.2-20$ (upper left to bottom right). Colors are retained across panels by the following rule: when a community is divided into multiple sub-communities, the sub-community that is the most connected to the original (``parent") community retains the color of the parent community; the other sub-communities are assigned new colors. The modularity for all of the panels is over 0.8.}
\end{figure*}

In order to further understand similarities between the mobility and communication networks, we performed a multi-scale analysis of community structure using a generalized modularity optimization algorithm that introduces a resolution parameter, $\gamma$ \cite{f4}. Smaller values of $\gamma$ identify progressively larger communities, and vice-versa. The multi-scale analyses of the mobility and communication networks are shown for some examples of $\gamma$ values in Figures \ref{fig:Mob} and \ref{fig:Men}, respectively. Partitions range from a single large module of the entire U.S. (top panels) down to urban scale partitions (bottom panels). Some states like Pennsylvania (PA) are split into multiple communities early in the process ($\gamma \approx 0.4$ in the mobility network), while other states like Texas (TX) emerge as single communities ($\gamma \approx 1$ in the mobility network) and internally fragment later in the process. These differences are directly associated with the internal structure of social ties and their geographical breakpoints, further explored in the Discussion (Section \ref{discussion}). In order to validate these partitions, we compared them with the communities detected by Infomap \cite{Bohlinmapequation}. This method finds the best partition based on the flow of information in a network. The comparison shows that the patterns obtained using Infomap are very similar to the ones obtained from the multi-scale modularity method at specific values of $\gamma$ (see Section S1.4 and Figure S5).

\begin{figure}[ht!]
\centering
\includegraphics[clip,trim={0cm 1cm 0cm 0cm},width=0.5
\textwidth ,angle =270]{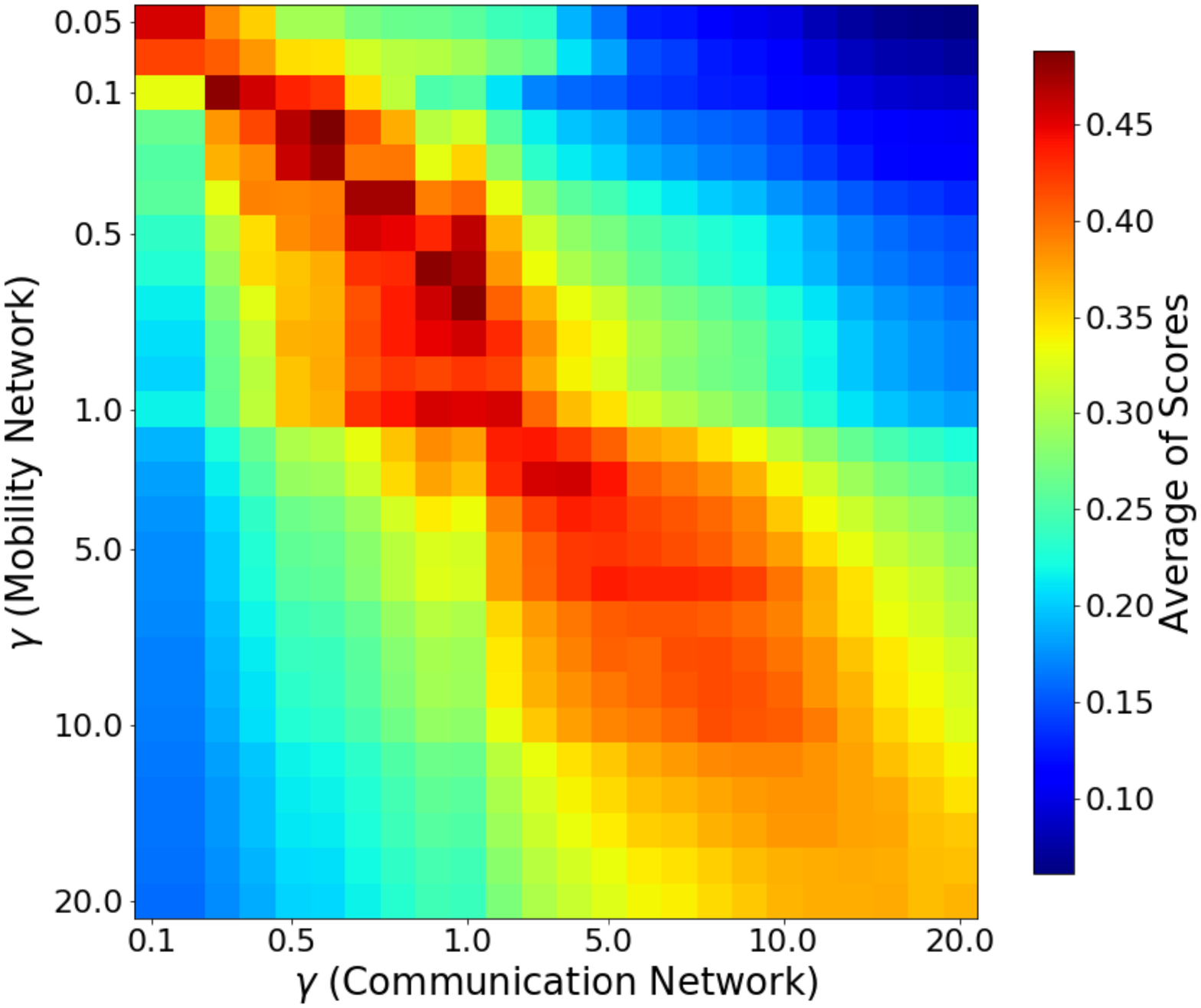}
\caption{\label{fig:sumscores} Similarity between the mobility and communication networks across multiple scales. Similarity is measured by the average of the Purity, Adjusted Rank, and Fowlkes-Mallows Indexes (color scale shown). Scale is defined by the different values of the resolution parameter $\gamma$ (horizontal and vertical axes). }
\end{figure}

We compare the partitions in both networks for different values of $\gamma$ by using three measures of cluster similarity: Purity \cite{f5}, Adjusted Rand Index \cite{f6} and Fowlkes-–Mallows Index \cite{f7}. These measures evaluate the overlap of partitions, with values ranging between 0 (no intersection) and 1 (perfect match). Figure \ref{fig:sumscores} shows a matrix whose rows and columns represent the partitions of the mobility and communication networks at different values of resolution ($\gamma$-mobility and $\gamma$-communication) and whose elements show the average of the three measures of similarity. The highest similarity between the two networks occurs at similar values of resolution (red diagonal), showing that the relative structure of these networks is consistent across scales. Additional comparisons between the two networks can be found in Section S1.5, including measures of degree centrality and edge weight (Figure S6) and an alluvial diagram (Figure S7). 

The consistency between the mobility and communication networks reveals that social spaces are not limited to the physical space. Instead, offline interactions seem to condition the structure of online communications. Moreover, the hierarchical multi-scale structure of these networks reveals that smaller communities with cohesive social ties, interactions, and associations belong to progressively larger ones. It may be expected that locations from the same community will have more similarity than locations from different communities.

\begin{figure}[ht!]
\centering
\includegraphics[clip,trim={0.5cm 2.8cm 0.5cm 1cm},width=0.9\textwidth,angle=270 ]{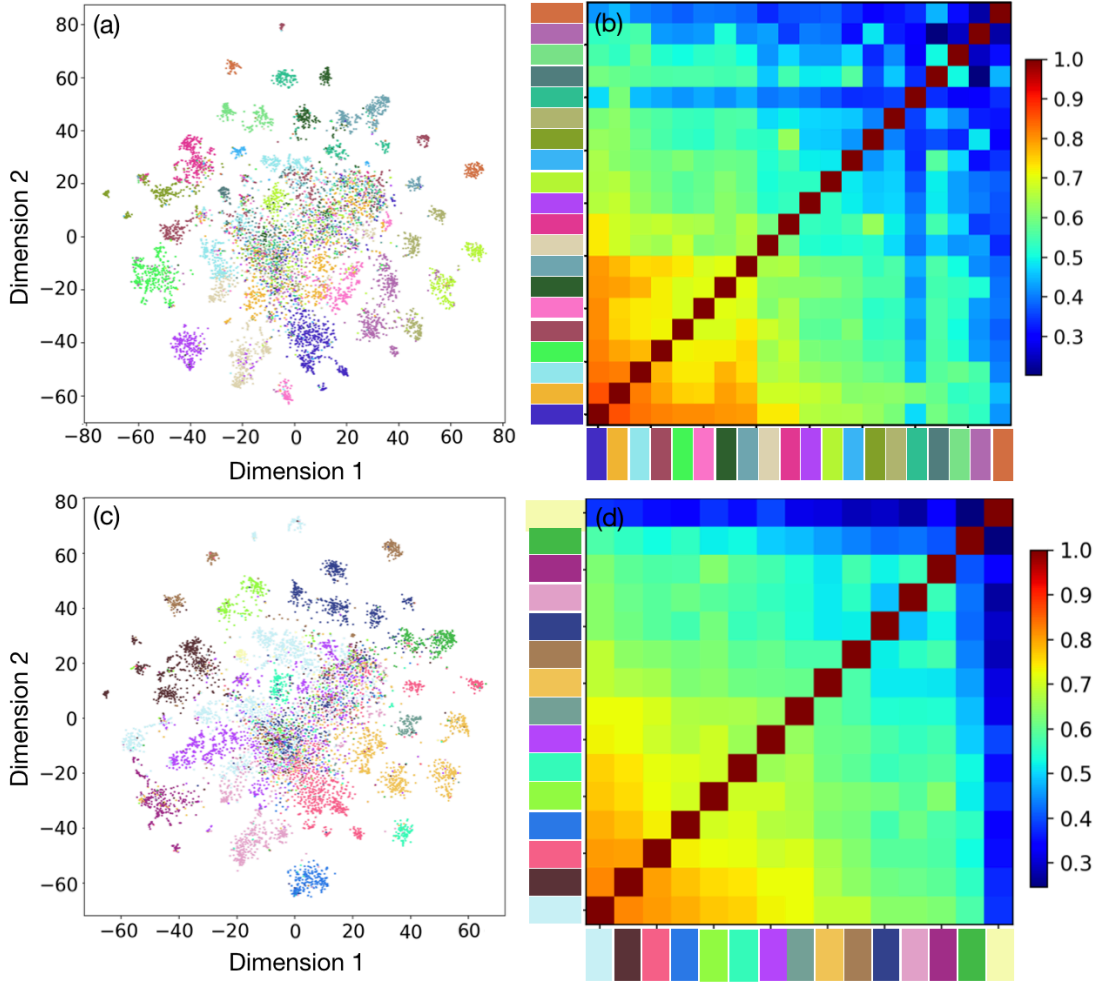}
\caption{\label{fig:hashtags} Dimensional reduction analysis of hashtag use by location and cosine similarity of communities based on hashtags. Panel (a) shows the results of t-SNE analysis on the first 100 components of PCA analysis of hashtags in locations of the mobility network. Panels (b) shows cosine similarity of hashtag use in the communities of the mobility network at $\gamma=1$. Panels (c) and (d) show the corresponding t-SNE result and cosine similarity for the communication network. Colors match those of the communities in Figures \ref{fig:Networks} and \ref{fig:NetworksS}, panel (b).}
\end{figure}

Locations from the same community show similarity in hashtag use and divergence with locations from different communities for either the mobility or communication networks (Figure 1(b) and 2(b)). Hashtags highlight specific, shared experiences and serve as markers of social interaction \cite{Zappavigna2015}. We compared hashtags for locations in the mobility and communication networks at $\gamma = 1$ using principal component analysis (PCA) followed by t-distributed stochastic neighbor embedding (t-SNE) analysis (Figure \ref{fig:hashtags} (a) and (c) for mobility and communication, respectively). See Section \ref{methods} for more information on the method. We colored each dot by location, matching the colors of the communities in Figures \ref{fig:Networks} and \ref{fig:NetworksS}, panel (b). A number of distinct colored clusters emerge, suggesting that hashtag use by location corresponds to communities of the mobility or communication networks. Some  clusters appear to separate into smaller clusters near to each other, representing sub-communities inside the communities. These patterns are statistically significant after randomizing locations ($p < 0.001$), detailed in Section S1.6 and Figure S8. To compare communities to each other based on hashtag use, we performed analysis of cosine similarity (Figure \ref{fig:hashtags} (b) and (d) for mobility and communication, respectively). Squares are colored from blue to red for increasing similarity (color bar, right). About half of community pairs have less than $50\%$ similarity, while the rest have $50-90\%$ similarity. Communities are distinct at some scales and form larger communities at higher scales.

\subsection{Model} 

We constructed a network growth model that combines aspects of network dynamics and human mobility in order to show the emergence of social fragmentation. Our model combines geographical distance gravity \cite{m2}, preferential attachment to allow creation of hubs (cities), and spatial growth to allow the growth of cities \cite{m1}. We begin with a lattice representing geographical locations, and grow connections among them simulating the way people travel. The probability of creating an edge between locations $i$ and $j$ in each time step is:
\begin{equation} \label{eq1}
P_{ij} \sim <k_{nn}>_{i}^{\nu}\frac{k_{j}^{\alpha}}{d_{ij}^{\beta}}
\end{equation}

where $i$ represents the origin of the interaction, $j$ indicates the destination, $<k_{nn}>_{i}$ indicates $i$'s  nearest neighbors' average degree, $k_j$ represents $j$'s degree, and $d_{ij}$ represents the distance between $i$ and $j$. The exponents $\alpha$, $\beta$ and $\nu$ control the effects of the preferential attachment mechanism, geographical distance gravity and spatial growth, respectively. The model reproduces the growth of geographical clusters similar to cities ($\nu$), their degree of attractiveness ($\alpha$) and the linkage between urban centers and surrounding areas, including neighboring cities ($\beta$). We introduce the preferential attachment mechanism to break the symmetry of spatial connections over time and the spatial growth mechanism to allow the city-like structures to grow.

Each location in the lattice has $4$ nearest neighbors, except for locations in corners and on edges, which have $2$ and $3$ neighbors, respectively. Simulations start with a random seed of three connected locations. Links are undirected and weighted to represent the iteration of links over time. Origins are picked randomly (independent from destinations) if their normalized value of $<k_{nn}>^{\nu}$ exceeds a random threshold. To allow all the locations in the lattice to participate in the dynamics, for the first $N$ time steps, we turn off the origin priority selection and let the system choose origins from a random order of locations, where $N$ represents the number of locations. The probability of selecting destinations is a combination of the preferential attachment mechanism and geographical distance gravity as shown in Equation \ref{eq1}. Thus, locations that are nearer to the origin location and have a higher degree have a higher probability to be chosen. Simulations continue until reaching a stable state in which communities form and do not change in number.

\begin{figure*}[p]
\centering
\includegraphics[clip,trim={0cm 7.5cm 4.3cm 2cm },width=0.58\textwidth]{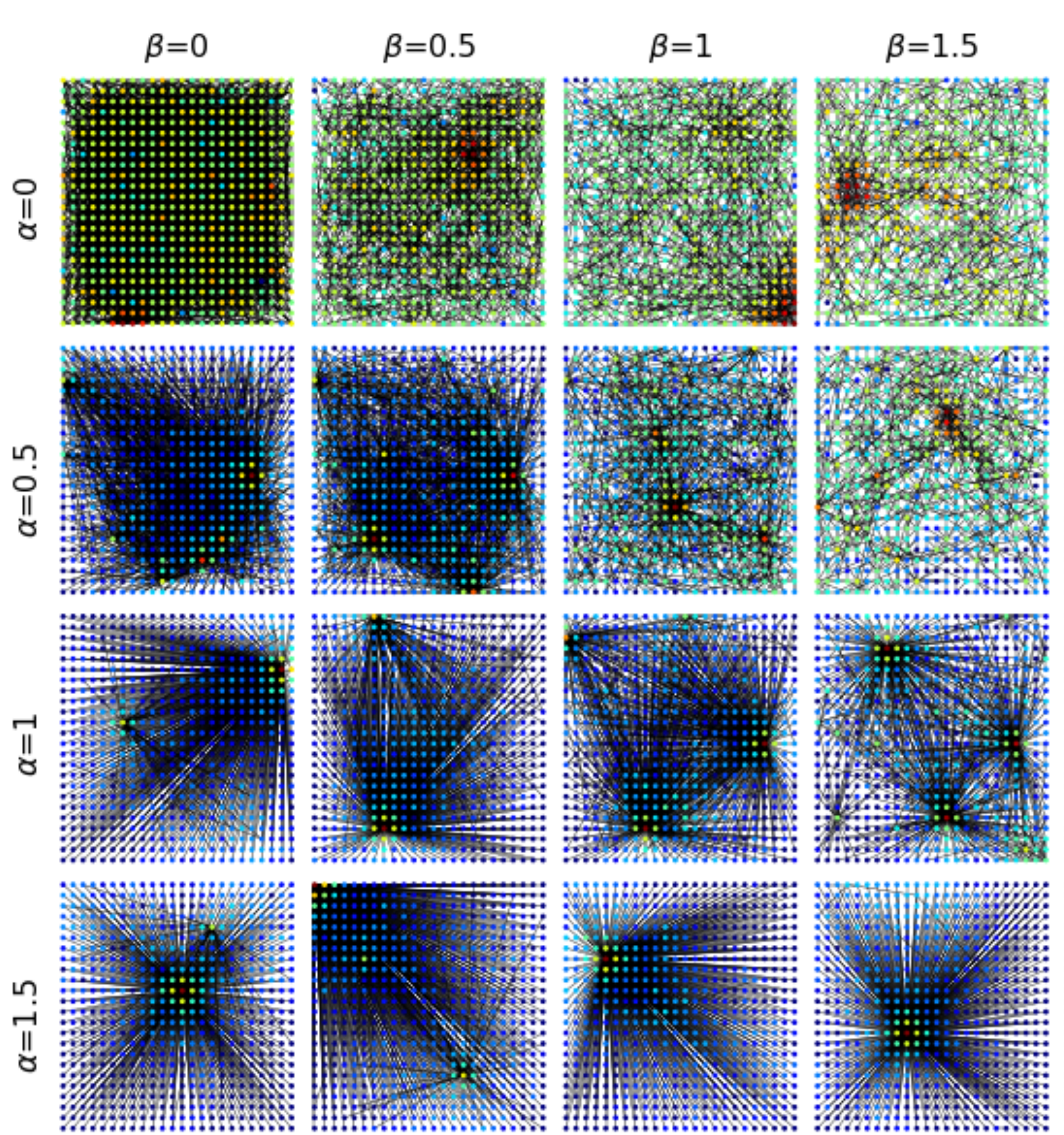}
\includegraphics[clip,trim={0cm 7.3cm 4.1cm 3cm },width=0.59\textwidth]{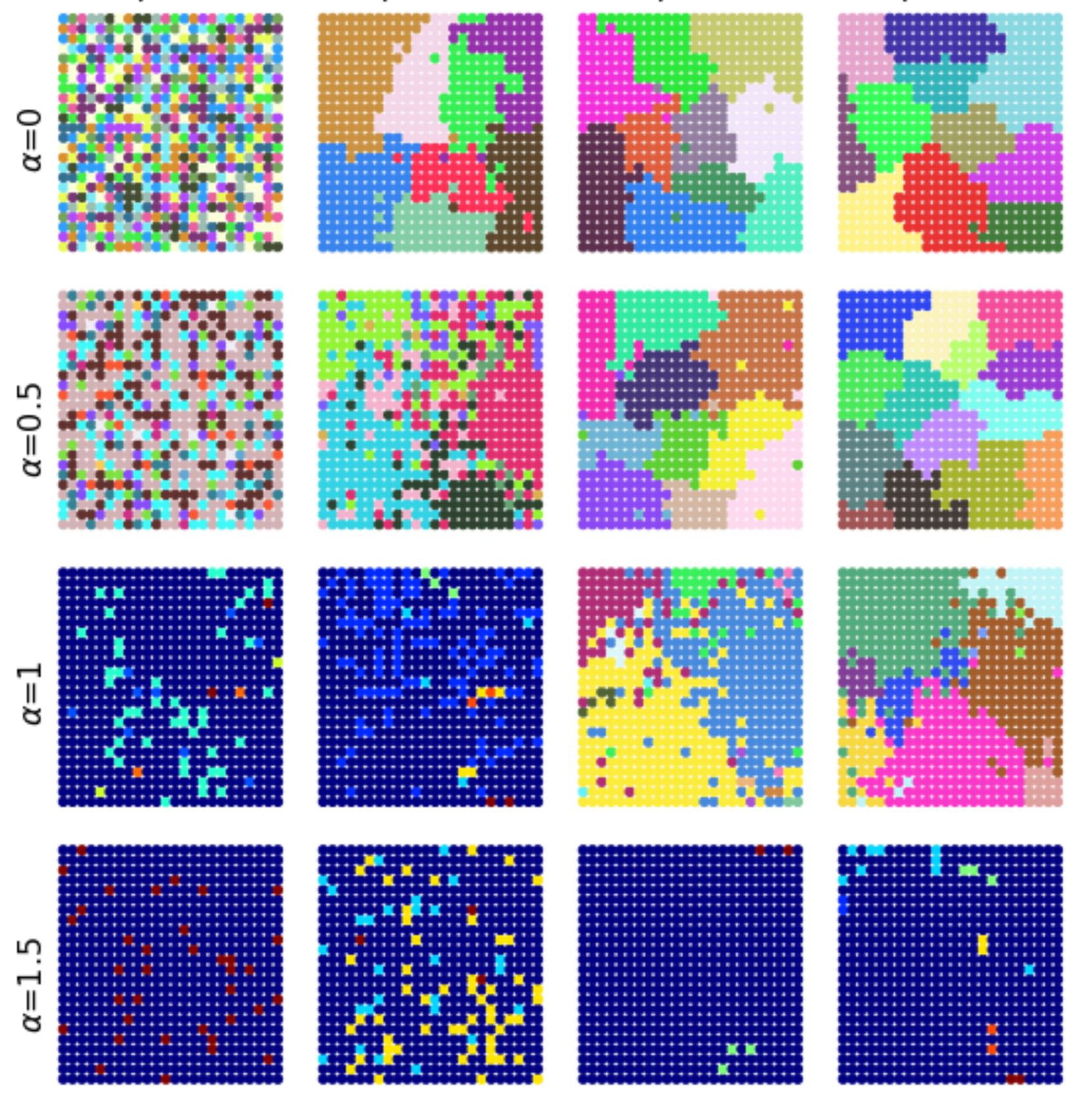}
\caption{\label{fig:model_1} Spatial degree distribution and modular structure for model simulations with different values for parameters $\alpha$ (preferential attachment) and $\beta$ (geographical distance gravity) and a fixed value of 0.1 for $\nu$ (spatial growth). Top panels show the spatial degree distribution (from weakly connected in blue to highly connected in red). Bottom panels show the modules of each graph, with each color identifying a single community. } 
\end{figure*}

Figure \ref{fig:model_1} shows the results of model simulations in terms of the spatial degree distribution (top panels) as well as modular structure (bottom panels) for different values of $\alpha$ (rows) and $\beta$ (columns) and a fixed value of $\nu=0.1$. If we do not include the effects of either preferential attachment ($\alpha=0$) or gravity ($\beta=0$), the destinations of edges are independently distributed among all nodes and the resulting communities have no spatial pattern. If $\alpha > \beta$, then a few hubs and one or two communities arise without significant geographic effects. Spatial fragmentation arises when the gravity mechanism is stronger than the preferential attachment  ($\beta>\alpha$), either without hubs ($\alpha=0$) or with hubs ($\alpha>0$). Increasing $\nu$ leads to more localized high-activity areas (cities), but this also destroys localized patches, leading to lower values of modularity. For additional results exploring variation of the spatial growth mechanism while keeping $\alpha$ and $\beta$ constant, see Section S1.7 and Figure S9.

\begin{figure}[ht!]
\centering
\includegraphics[clip,trim={3cm 4.5cm 1cm 4.5cm },width=0.55\textwidth]{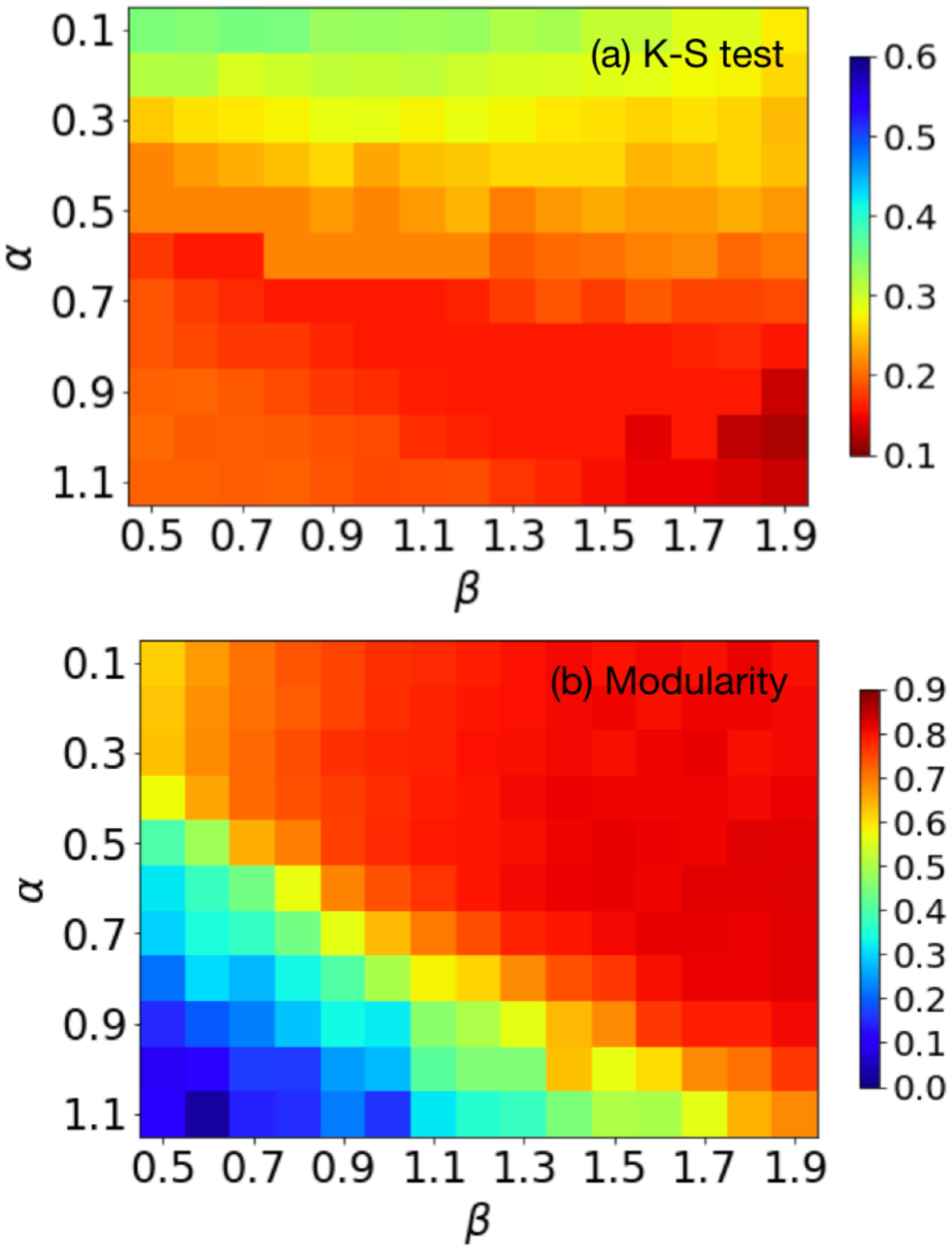} 
\caption{\label{fig:model_2} Kolmogorov-Smirnov score and modularity for simulations with varying model parameter, $\alpha$ and $\beta$, and fixed resolution parameter $\nu=0.1$. (a) Colors indicate the Kolmogorov-Smirnov (K-S) score, with lower scores (red) indicating similarity between the degree distributions of the model and the mobility network. (b) Colors indicate network modularity. Modularity is highest at around $0.8$ (dark red), similar to the actual modularity for the mobility network ($0.83$).}
\end{figure}

We validated the model results against Twitter data by first testing whether the degree distributions from both sources are drawn from the same distribution and second comparing the modularity values. For each set of parameters, we created $20$ model realizations and analyzed their statistical behavior. We applied the Kolmogorov-Smirnov statistical test (K-S) to compare the average degree distribution from the model realizations to that of the mobility network, and similarly for the communication network. Figure \ref{fig:model_2} (a) shows the values of the test results for different values of $\alpha$ and $\beta$ (rows and columns of the matrix) and $\nu=0.1$. Lower K-S values (red) indicate more similarity, and higher K-S values (blue) indicate less similarity. The average modularity values for the simulations in Figure \ref{fig:model_2} (a) are shown in Figure \ref{fig:model_2} (b), ranging from 0 (no modular structure) to 1 (high modular structure). We find that $\alpha=0.9$, $\beta=1.5$, and $\nu=0.1$ give a good fit between simulations and observed data. Results for the K-S statistic with variation of all three parameters are shown in Section S1.7 and Figures S10 and S11. 

\section{Discussion}\label{discussion} 
Understanding the structure and dynamics of groups is an essential aspect of understanding social interactions generally. The functioning of human societies arises not only from the activities of individuals but also from their interaction and integration by means of social ties. We analyzed the structure of social ties in the U.S. using Twitter data and found multi-scale, self-organized fragments that span from urban up to national scales for mobility, communication, and hashtag use. Our results show that the structures emerging from these different types of interaction are highly consistent, revealing that social ties couple the integration and separation of groups in both physical and virtual spaces. Despite potential biases in Twitter samples \cite{Duggan2015, Smith2018}, the similarity of the detected communities between mobility and communication networks shows that the networks reveal the underlying social structure.
 
We also constructed a model of network growth that is consistent with the statistical property of emergence of the observed patterns from the Twitter data. Our model shows that social fragmentation may result from short-distance interactions, in support of hierarchical models of social network formation \cite{Watts1302}. However, this mechanism alone does not explain the emergence of highly connected places such as cities. We model the emergence of cities using preferential attachment and spatial growth mechanisms, which increase heterogeneity in degree distribution but may destroy spatial fragmentation if cities grow large enough. Other generative models can also create fat-tails and power-law behaviors. For example, the emergence of city centers can also be modeled as  processes of optimization of social interactions and information flows or as outcomes of multiplicative growth mechanisms \cite{Mitzenmacher2004}.

The gravity model \cite{10.2307/2087063,m3} describes how the strength of mobility between two locations is directly related to the population density of those locations and inversely related to the distance between the locations, each of which is a power-law relation. Reported values for the scaling exponents vary in the range $0.5-2.0$ depending on the system \cite{m11,m5,m12,m9,m7,Junjun2017}. Thus, the gravity model predicts that cities with higher population densities attract higher mobility. However, an important limitation is that the gravity model may overestimate mobility from a low density population to a high density population, limiting its applicability over wide geographic areas \cite{m2}. Furthermore, the gravity model does not allow for cluster growth or changes in the population of locations. We overcome these limitations by creating a model incorporating geographical distance gravity with preferential attachment and spatial growth.

The formation of groups and their interactions are intimately related to the formation of individual identity through self-identification and adoption of group norms and narratives. Thus, while individual identities are highly complex and unique, there are shared patterns among members of self-associating groups. These common patterns define the group identity, which may involve linguistic, cultural, economic, opinion or interest differences from other groups. To investigate divergence of shared social experience, we analyzed hashtag use by location. Hashtags are a means of discussing shared experiences and ideas, aspects of group formation. Our analysis demonstrated that many of the communities from the mobility or communication networks have also distinct hashtag use. This suggests that the communities shared experiences also diverge from other communities.

When we further examine the mobility and communication networks at different scales (Figures \ref{fig:Mob} and \ref{fig:Men}), we observe that many communities follow state lines, but a few do not, suggesting other forces driving community formation. The large metro area around St. Louis, MO creates a community that spills across the Mississippi River and thus the Missouri (MO)-Illinois (IL) state line (mobility network: $\gamma \approx 0.7-20$, communication network: $\gamma \approx 1$). Eastern and western Pennsylvania (PA) splits into two communities, roughly along the Appalachian mountain boundary (mobility network: $\gamma \approx 0.4-1$, communication network: $\gamma \approx 0.6-1$). California (CA) splits into northern and southern communities (mobility network: $\gamma \approx 0.7-1$, communication network: $\gamma \approx 2$), following a known cultural and economic divide \cite{DiLeo1980, Bucholtz2007}. The area of eastern Idaho (ID) combines with Utah (UT) (mobility network: $\gamma \approx 0.3-0.7$, communication network: $\gamma \approx 0.7-2$), corresponding to the area of historical Mormon settlement \cite{Meinig1965}. Geographers have proposed that many cultural, political, and religious divisions trace back to the original settlers in each area \cite{Zelinsky1973}, such that America can be divided into corresponding cultural regions or ``nations" \cite{Woodard2011}, which has largely been supported by recent genetic studies of the U.S. population \cite{Han2017}. Our observations also support that the communities we observe reflect geographical, cultural, and economic forces that can supersede administrative boundaries in some locations, although state boundaries remain an important factor in social interactions.

Recent trends seem to be accelerating the forces of community formation and divergence we observe. These forces include economic shifts, political polarization, and the rise of social media. Analyses of work commutes have supported the rise of ``megaregions,'' interconnected labor markets with large cities as hubs, reminiscent of the communities we observe \cite{DashNelson2016}. Migrations from one megaregion to another may be motivated by economics, such as the migration over the last decade from the Northeast toward the mountain West and Southwest, which have offered better job prospects and lower housing prices \cite{Gray2018}. In addition to economic movements, an increase in political self-sorting behavior has been observed, with people physically moving nearer to like-minded individuals \cite{Johnston2016}. The percentage of people who identify as ``consistently" liberal or conservative has doubled to over 20\% in the past two decades, and these individuals express preferences to live near to, be close friends with, and marry those of the same political persuasion \cite{ResearchC2014}. Social media may be exacerbating this polarization, creating spaces in which users interact with like-minded individuals and ignore opposing opinions \cite{I28}. Future work will need to examine how patterns of group formation change and whether cultural, political, or economic factors drive this polarization.

Moving forward, there are at least two strategies for policymakers seeking to address social fragmentation in the U.S. One is to fight social fragmentation by promoting intergroup connection and uniformity in society. The other is to recognize that social fragmentation is present and to incorporate it into policy decisions. This means adopting a policy of localism, which involves tailoring policy approaches to each specific area and fostering participation from local political groups \cite{Ercan2013}. Our analysis suggests that division into two political groups (e.g., Republican and Democrat) is not sufficient in the U.S. today and that sub-groups may require partial local autonomy to address the multi-scale divisions present in society.

\section{Conclusion}
In summary, we have used geo-located Twitter data to generate networks of U.S. mobility, communication, and hashtag use and to explore how networks fragment at multiple scales. We also developed a model of network growth that incorporates the properties of geographical distance gravity, preferential attachment, and spatial growth and successfully replicates statistical properties of the social fragmentation patterns observed in the data. Overall, our analysis demonstrates there are many boundaries along which fragmentation of U.S. society may be taking place. Moreover, this fragmentation represents a multi-factorial and dynamic process that is ongoing. It is an important question how social fragmentation at multiple levels will affect the stability and dynamism of U.S. society in the future.

\section*{Acknowledgments}{We thank Irwin Epstein and William Glenney for feedback and Matthew Hardcastle for proofreading the manuscript.}

\section*{Author contributions:} Conceptualization: L.H., A.J.M., Y.B-Y. Data curation: A.J.M. Formal analysis: L.H., A.J.M. Supervision: L.H., A.J.M., Y.B-Y. Writing, original draft: L.H., R.A.R., A.J.M. Writing, review, and editing: L.H., R.A.R., A.J.M, Y.B-Y. Funding acquisition: Y.B-Y.

\section*{Funding:} New England Complex Systems Institute

\section*{Competing Interests:} The authors declare no competing interests.

\section*{Data and material availability:} Data are available at: https://necsi.edu/fragmentation-data

\pagebreak

\section*{Figure Legends}
\textbf{Figure \ref{fig:Networks}.} Structure and fragmentation patterns of the network associated with human mobility. (a) Spatial degree centrality of the mobility network. Colors indicate the amount of people traveling at each location, measured by the logarithm of the degree centrality of each node (scale inset). The mobility network was used to generate communities using modularity optimization, with distinct colors indicating (b) 20 patches that can be visually associated to states or regions and (c) 206 smaller sub-communities within the communities of panel (b) that can be visually associated to urban centers.

\textbf{Figure \ref{fig:NetworksS}.} Structure and fragmentation patterns of the network associated with human communication. (a) Spatial degree centrality of the communication network. Colors indicate the amount of communication at each location, measured by  the logarithm of the degree centrality of each node (scale inset). The communication network was used to generate communities using modularity optimization, with distinct colors indicating (b) 15 patches that can be visually associated to states or regions and (c) 168 smaller sub-communities within the communities of panel (b).

\textbf{Figure \ref{fig:Sim}.} Similarity of communities in the communication and mobility networks. Matrix of the regional communities for the communication network ($y$\mbox{-}axis, $n=15$) and mobility network ($x$\mbox{-}axis, $n=20$), ordered by decreasing overlap between communities. Cell colors represent the number of nodes overlapping between the two networks in each community, normalized by the size of the communities per row (scale inset), with no overlap indicated in white. 

\textbf{Figure \ref{fig:Mob}.} Multi-scale decomposition of the mobility network. Colors indicate geographical patches detected in the mobility network for values of the resolution parameter $\gamma$ varied from $0.05-20$ (upper left to bottom right). $M$ stands for modularity.

\textbf{Figure \ref{fig:Men}.} Multi-scale decomposition of the communication network. Colors indicate geographical patches detected in the communication network for values of the resolution parameter $\gamma$ varied from $0.05-20$ (upper left to bottom right). $M$ stands for modularity.

\textbf{Figure \ref{fig:sumscores}.} Similarity between the mobility and communication networks across multiple scales. Similarity is measured by the average of the Purity, Adjusted Rank, and Fowlkes-Mallows Indexes (color scale shown). Scale is defined by the different values of the resolution parameter $\gamma$ (horizontal and vertical axes).

\textbf{Figure \ref{fig:hashtags}.} t-SNE analysis of hashtag use by location and cosine similarity of communities based on hashtags. Panel (a) shows the results of t-SNE analysis on the first 100 components of PCA analysis of hashtags in locations of the mobility network. Panels (b) shows cosine similarity of hashtag use in the communities of the mobility network at $\gamma=1$. Panels (c) and (d) show the corresponding t-SNE result and cosine similarity for the communication network. Colors match those of the communities in Figures \ref{fig:Networks} and \ref{fig:NetworksS}, panel (b).

\textbf{Figure \ref{fig:model_1}.} Spatial degree distribution and modular structure for model simulations with different parameter values  $\alpha$ and $\beta$ and a fixed value of $\nu=0.1$ (see text for parameter definitions). Top panels show the spatial degree distribution (from weakly connected in blue to highly connected in red). Bottom panels show the modules of each graph; each color identifies a single community. 

\textbf{Figure \ref{fig:model_2}.} Kolmogorov-Smirnov score and modularity for simulations with varying model parameter, $\alpha$ and $\beta$, and fixed resolution parameter $\nu=0.1$. (a) Colors indicate the Kolmogorov-Smirnov (K-S) score, with lower scores (red) indicating similarity between the degree distributions of the model and the mobility network. (b) Colors indicate network modularity. Modularity is highest at around $0.8$ (dark red), similar to the actual modularity for the mobility network ($0.83$). 
\raggedbottom 

\end{document}